\documentclass[review]{elsarticle}
\makeatletter
\def\ps@pprintTitle{
	\let\@oddhead\@empty
	\let\@evenhead\@empty
	\def\@oddfoot{\reset@font\hfil\thepage\hfil}
	\let\@evenfoot\@oddfoot
}
\makeatother
\biboptions{comma,sort&compress}

\usepackage{graphicx}
\usepackage{amsmath}
\usepackage{amssymb}
\usepackage{here}
\usepackage{cuted}

\usepackage[dvips]{epsfig}
\def\nin{\noindent}
\def\beq{\begin{equation}}
\def\eeq{\end{equation}}
\def\bea{\begin{eqnarray}}
\def\eea{\end{eqnarray}}
\def\nnb{\nonumber}
\def\la{\langle}
\def\ra{\rangle}

\usepackage{graphicx}
\usepackage{here}
\def\beq{\begin{equation}}
\def\eeq{\end{equation}}
\def\bea{\begin{eqnarray}}
\def\eea{\end{eqnarray}}
\def\bq{\begin{quote}}
\def\eq{\end{quote}}
\def\ve{\vert}
\def\nnb{\nonumber}

\def\nnb{\nonumber}
\def\la{\langle}
\def\ra{\rangle}
\def\nin{\noindent}
\def\ba{\begin{array}}
\def\ea{\end{array}}

\def\als{\alpha_s}

\def\gg2{ \la\alpha_s G^2 \ra}
\def\gg3{g^3f_{abc}\la G^aG^bG^c \ra}
\def\ggg4{\la\als^2G^4\ra}

\def\beq{\begin{equation}}
\def\enq{\end{equation}}
\def\beqa{\begin{eqnarray}}
\def\enqa{\end{eqnarray}}
\def\nnb{\nonumber}

\def\MeV{\nobreak\,\mbox{MeV}}
\def\GeV{\nobreak\,\mbox{GeV}}
\def\keV{\nobreak\,\mbox{keV}}

\newcommand{\rag}{\rangle}
\newcommand{\lag}{\langle}

\def\gg{\lag g^{2}_{s} G^2 \rag}
\def\ggg{\lag g^{3}_{s}G^3\rag}
\begin{document}
%\markboth{Stephan Narison, Montpellier (FR)}{ }
\begin{frontmatter}

\title{Light scalar quarkonia from QCD Laplace sum rule at higher order\tnoteref{text1}}

\author[label1]{R. M. Albuquerque}
\ead{raphael.albuquerque@uerj.br}
\address[label1]{Faculty of Technology,Rio de Janeiro State University (FAT,UERJ), Brazil}
\author[label2,label3]{S. Narison}
%\fntext[fn1]{ICTP-Trieste consultant for Madagascar}
\ead{snarison@yahoo.fr}
\address[label2]{Laboratoire
Univers et Particules de Montpellier (LUPM), CNRS-IN2P3, Case 070, Place Eug\`ene
Bataillon, 34095 - Montpellier, France}
\address[label3]{Institute of High-Energy Physics of Madagascar (iHEPMAD), University of Antananarivo, Antananarivo 101, Madagascar}

%\author[label3]{A. Rabemananjara}
%\ead{achrisrab@gmail.com} 

\author[label3]{D. Rabetiarivony\fnref{fn2}}
\fntext[fn2]{Speaker}
\ead{rd.bidds@gmail.com} 

%\author[label3]{G.~Randriamanatrika}
%\ead{artesgaetan@gmail.com}

\tnotetext[text1]{Talk given at QCD25 International Conference (30 june -- 04 july 2025, Montpellier--FR)}

\pagestyle{myheadings}
\markright{ }

\date{\today}
\begin{abstract}
%\noindent
We review our estimations on the light scalar $\bar{q}q$, $(\bar{q}q')(\bar{q'}q)$ and $\overline{qq'}qq'$ ($q,q'\equiv u,d,s$) states from relativistic Laplace sum rule (LSR) within stability criteria and including higher order perturbative (PT) corrections up to the (estimated) N5LO. We evaluate the QCD spectral functions at Lowest Order (LO) of PT QCD and up to the $D=6$ dimension of quark and gluon condensates. Using stability criteria and the constraint: Pole contribution is larger than the QCD continuum one ($R_{P/C}\geqslant 1$) our results exclude an on-shell mass around $(500-600)~\MeV$ obtained for values of the QCD continuum threshold $t_c \leqslant(1\sim 1.5)~\GeV^2$. The complete results for the different scalar states are given in Tables\,\ref{tab:resqqmol} to \ref{tab:resva}. We conclude from the complete analysis that the assignement of the nature of the scalar mesons is not crystal clear and needs further studies. 

\end{abstract}
\end{frontmatter}
%%%%%%%%%%%%%%%%%%%%%%%%%%%%%%%%%%
%\newpage
%%%%%%%%%%%%%%%%%%%%%%%%%%%%%%%%%%%%%%%%%%%
\section{Introduction}
%%%%%%%%%%%%%%%%%%%%%%%%%%%%%%%%%%%%%%%%%%%
In this talk, we summarizes the results of our works in Ref.\,\cite{ANRls}.

\nin Despite many theoretical and experimental efforts, the true nature of the light scalar mesons remains still not clear. Several configurations have been considered for these states. Among different theoretical interpretations of the light scalar mesons we have: ordinary mesons $\bar{q}q$\, \cite{SNB4,SNA,MNP,NPRT,SND,RRY,SNB5,BSN} and gluonia\,\cite{SNV,SNB4,SNA}, \,four-quark states and $\pi^+\pi^-,\,K^+K^-,\,K\pi,\cdots$ molecules\,\cite{JAFFE,ADS,IW,SNB2,SNB5,LP,BNNB,CHZ,CS}. In this paper, we improve the analysis of the estimations of the masses and couplings of ordinary $\bar{q}q$ and exotic tetraquark configurations using the Laplace sum rule\,\cite{BELL,BELLa,BNR,BERT,NEUF,SNR,SNB1,SNB2} version of QCD spectral sum rules (QSSR)\footnote{For reviews, see \cite{SVZa,Za,SNB1,SNB2,SNB3,CK,YND,PAS,RRY,IOFF,DOSCH,SNqcd22} and references therein} and present new results for the $\pi^+\pi^-,\,K^+K^-,\,K\pi$ and $\eta\pi^0$ molecules.
%%%%%%%%%%%%%%%%%%%%%%%%%%%%%%%%%%%%%%%%%%%
\section{The Laplace sum rule}
%%%%%%%%%%%%%%%%%%%%%%%%%%%%%%%%%%%%%%%%%%%
We shall work with the inverse  Laplace transform sum rule and their ratios for extracting the decay constant and the mass:
\bea
\hspace*{-0.2cm} {\cal L}^c_n(\tau,\mu)&\equiv & \int_{t_>}^{t_c} dt~t^n~e^{-t\tau}\frac{1}{\pi} \mbox{Im}~\psi_{S}(t,\mu)~;\nnb \\
 {\cal R}^c_{10}(\tau)&\equiv &\frac{{\cal L}^c_{1}} {{\cal L}^c_0},
\label{eq:lsr}
\eea
The spectral function $\mbox{Im}~\psi_S(t,\mu)$ can be evaluated from the two-point correlator:
\bea
\hspace*{-0.6cm}
\psi_S(q^2)\hspace*{-0.2cm}&=& \hspace*{-0.2cm} i \int \hspace*{-0.1cm} d^4 x\, e^{i q x}\lag 0 \ve \mathcal{T} \mathcal{O}_S(x)(\mathcal{O}_S(0))^{\dag} \ve 0 \rag,
%&\equiv &\hspace*{-0.2cm} -\left( g^{\mu\nu}-\frac{q^{\mu}q^{\nu}}{q^2}\right)\Pi^{(1)}_{\mathcal{H}}(q^2)+\frac{q^{\mu}q^{\nu}}{q^2}\Pi^{(0)}_{\mathcal{H}}(q^2),
\eea
where $\mathcal{O}_S(x)$ are the local hadronic operators describing the $\bar{q}q$ mesons, $(\bar{q}q')(\bar{q}'q)$ molecules and $(\overline{qq'})(qq')$ tetraquark states; $t_>$ is the quark threshold and $\tau$ is the LSR variable; $\mu$ is an arbitrary perturbative (PT) subtraction constant which is equal to $1/\sqrt{\tau}$ as we shall work with a renormalization group resumed PT series; $t_c$ is the threshold of the "QCD continuum".
%Contrary to some intuitive claims in the literature, $\sqrt{t_c}$ does not necessarily coincide with the mass of the first radial excitation but can be higher as the QCD continuum smears all higher state contributions.
%\nin The QCD expressions of the leading order (LO) spectral functions up to dimension 6 condensates, details about the FNLO PT corrections and the different QCD input parameters are given in Ref.\,\cite{SNqcd22,PICH,NPIV,HAGIWARA,RRY}.
%%%%%%%%%%%%%%%%%%%%%%%%%%%%%%%%%%%%%%%%%%%
\section{Stability criteria}
%%%%%%%%%%%%%%%%%%%%%%%%%%%%%%%%%%%%%%%%%%%
Like in our previous works\,\cite{LNSR,ANRls,ANRTm,ANR21,ANR22p,ANR22pa,ANRR1,ANRR1a,ANRR2,ANR1,NR1,SNX1,SNX2,SU3},  one considers that the optimal values of the masses and couplings should be independent (minimum sensitivity) on the variation of the external parameters $\tau$ and $t_c$. To check the conservative region of $(\tau,t_c)$, we shall also implement the rigorous condition:
\beq
\hspace*{-0.2cm}
R_{P/C}\equiv \frac{\mbox{Lowest Pole}}{\mbox{QCD continuum}}= \frac{\int^{tc}_{t_>}dt e^{-t\tau}\mbox{Im}\psi_S (t)}{\int^{\infty}_{t_c}dt e^{-t\tau}\mbox{Im}\psi_S (t)}\geqslant 1.
\eeq
Set of $(\tau,t_c)$ stability not satisfying this requirement will be rejected. 
%%%%%%%%%%%%%%%%%%%%%%%%%%%%%%%%%%%%%%%%%%%
%%%%%%%%%%%%%%%%%%%%%%%%%%%%%%%%%%%%%%%%%%%
\section{The ordinary $\bar{q}q$ states}
%%%%%%%%%%%%%%%%%%%%%%%%%%%%%%%%%%%%%%%%%%%
%%%%%%%%%%%%%%%%%%%%%%%%%%%%%%%%%%%%%%%%%%%
We shall be concerned with the quark operators:
\bea
J_2&=&\frac{m_q}{\sqrt{2}}(\bar{u}u+\bar{d}d),\nnb\\
J_{\bar{u}s}&\equiv& \partial_{\mu}V^{\mu}_{\bar{u}s}=(m_u-m_s)\bar{u}s,\nnb \\
J_3&=&m_s\bar{s}s,
\eea
where $m_q=(m_u+m_d)/2$.\\
We improve the estimation in \cite{SND,NPRT,RRY,SNB2,SNB5,BSN} by adding ${\cal O}(\alpha_s^5)$ higher order PT corrections\,\cite{ANRls} and LO non-perturbative contributions up to dimension $D=6$\,\cite{SNB1,NPRT,JM,SNB6}. The optimal values of the masses and couplings are extracted at N3LO assuming a violation of the factorization assumption. We consider that the error due to the truncation of the PT series comes from the sum N4LO$\oplus$N5LO (estimated by assuming that the coefficients of the series have a geometrical growth behaviour\,\cite{SNZ}). The final results with different sources of the errors are quoted in Table\,\ref{tab:resqqmol}.

%The stability region is delimited in the range $(t_c,\tau)=(2.0, 0.9) \sim (4.5, 1.5)\,(\GeV^2,\GeV^{-2})$ where, we obtain:
%\beq
%\hspace*{-0.1cm}
%M_{\bar{q}q} = 1246(96)\MeV,~f_{\bar{q}q}/m_q (\tau)= 274(43)\times 10^{-3}.
%\label{eq:fresqq}
%\eeq
%%%%%%%%%%%%%%%%%%%%%%%%%%%%%%%%%%%%%%%%%%%
%%%%%%%%%%%%%%%%%%%%%%%%%%%%%%%%%%%%%%%%%%%
\section{$\pi^+\pi^-\,,K^+K^-\,,K^+\pi^-\,,\eta\pi^0$ states}
%%%%%%%%%%%%%%%%%%%%%%%%%%%%%%%%%%%%%%%%%%%
%%%%%%%%%%%%%%%%%%%%%%%%%%%%%%%%%%%%%%%%%%%
We shall work with the following interpolating currents:
\bea
{\cal O}_{\pi^+\pi^-}(x)&=&(\bar{d}\,i\gamma_5\, u) \otimes (\bar{u}\, i\gamma_5\,d)(x),\nnb\\
{\cal O}_{K^+K^-}(x)&=&(\bar{s}\,i\gamma_5\, u)\otimes(\bar{u}\,i\gamma_5\, s)(x),\nnb\\
{\cal O}_{K^+\pi^-}(x)&=&(\bar{s}\,i\gamma_5\, u)\otimes(\bar{u}\,i\gamma_5\, d)(x),\nnb\\
{\cal O}_{\eta\pi^0}(x)&=&\hspace*{-0.3cm}\frac{1}{\sqrt{6}}\left[(\bar{u}\,i\gamma_5 u)+(\bar{d}\,i\gamma_5 d)-2(\bar{s}\,i\gamma_5 s)\right]\nnb\\
&&\otimes \frac{1}{\sqrt{2}}\left[(\bar{u}\,i\gamma_5 u)+(\bar{d}\,i\gamma_5 d)\right](x).
\eea
\nin We do not consider the scalar $\bar{q}q'$ current which cannot contribute to leading order to the decay $\sigma\rightarrow \pi^+\pi^-$. As the analysis will be performed using the same techniques, we shall illustrate it in the case of di-pion molecule state. The results are compiled in Table \ref{tab:resqqmol}.
%%%%%%%%%%%%%%%%%%%%%%%%%%%%%%%%%%%%%%%%%%%
%%%%%%%%%%%%%%%%%%%%%%%%%%%%%%%%%%%%%%%%%%%
\subsection*{$\bullet$ Mass and coupling of $\pi^+\pi^-$}
\begin{figure}[hbt]
\begin{center}
\includegraphics[width=7.0cm,height=4.3cm]{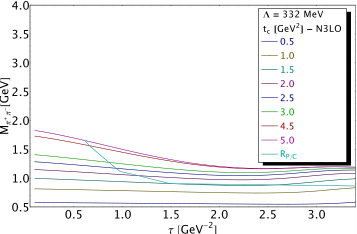}
\includegraphics[width=7.5cm,height=4.3cm]{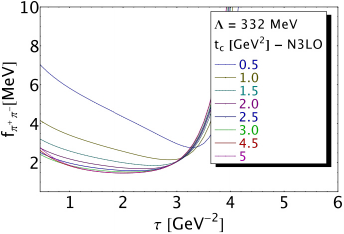}
%\vspace*{-0.5cm}
\caption{\footnotesize The mass and coupling of $\pi^+\pi^-$ as a function of $\tau$ at N3LO for \# values of $t_c$.} 
\label{fig:pp-n3lo}
\end{center}
\end{figure}

\begin{figure}[hbt]
\begin{center}
\includegraphics[width=7.0cm,height=4.3cm]{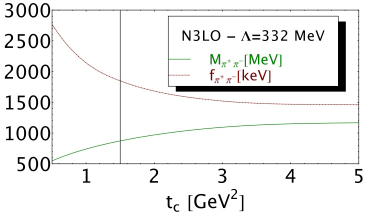}
%\vspace*{-0.5cm}
\caption{\footnotesize $t_c$ behaviour of the optimal results of $\pi^+\pi^-$.} 
\label{fig:pp-rpc}
\end{center}
\end{figure}

%%%%%%%%%%%%%%%%%%%%%%%%%%%%%%%%%%%%%%%%%%%
%%%%%%%%%%%%%%%%%%%%%%%%%%%%%%%%%%%%%%%%%%%
\nin We estimate the mass and the coupling respectively from the ratio of moments ${\cal R}_{10}$ and the low moment ${\cal L}_{0}$. The behaviour of the mass and the one of the coupling at N3LO of PT series are shown in Fig.\,\ref{fig:pp-n3lo}. We show in Fig.\,\ref{fig:pp-rpc} the $t_c$ behaviour of the optimal results. The vertical line ($t_c=1.5\GeV^2$) is the minimum value of $t_c$ allowed by $R_{P/C}\geqslant 1$ condition. The stability region is delimited in the range $t_c=(1.5 \sim 4.5)\,\GeV^2$ and  $\tau=(2.3 \sim 2.5)\,\GeV^{-2}$.
%%%%%%%%%%%%%%%%%%%%%%%%%%%%%%%%%%%%%%%%%
%%%%%%%%%%%%%%%%%%%%%%%%%%%%%%%%%%%%%%%%%
%%%%%%%%%%%%%%%%%%%%%%%%%%%%%%%%%%%%%%%%%
%%%%%%%%%%%%%%%%%%%%%%%%%%%%%%%%%%%%%%%%%
%%%%%%%%%%%%%%%%%%%%%%%%%%%%%%%%%%%%%%%%%%%
%%%%%%%%%%%%%%%%%%%%%%%%%%%%%%%%%%%%%%%%%%%
\subsection*{$\bullet$ Truncation of the PT series}
\begin{figure}[hbt]
\begin{center}
\includegraphics[width=7.0cm,height=4.3cm]{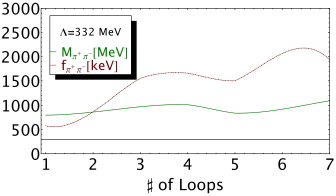}
%\vspace*{-0.5cm}
\caption{\footnotesize Behaviour of the optimal results versus the truncation of the PT series for $\pi^+\pi^-$ molecule. We take $t_c=2.31\GeV^2$ which corresponds to the central value of the mass and the one of the coupling. $\#7$ in the loops axis corresponds to the tachyon gluon mass contribution.} 
\label{fig:pp-pt}
\end{center}
\end{figure}

%%%%%%%%%%%%%%%%%%%%%%%%%%%%%%%%%%%%%%%%%%%
%%%%%%%%%%%%%%%%%%%%%%%%%%%%%%%%%%%%%%%%%%%
\nin We study in Fig.\,\ref{fig:pp-pt} the behaviour of the results versus the truncation of the PT series. One can notice that a stability is obtained for N2LO-N3LO which we consider as our optimal results. Like in the previous case of $\bar{q}q$ state, we consider as sources of the error due to the truncation of the PT series the sum N4LO$\oplus$N5LO.
%%%%%%%%%%%%%%%%%%%%%%%%%%%%%%%%%%%%%%%%%
%%%%%%%%%%%%%%%%%%%%%%%%%%%%%%%%%%%%%%%%%
%%%%%%%%%%%%%%%%%%%%%%%%%%%%%%%%%%%%%%%%%%%
%%%%%%%%%%%%%%%%%%%%%%%%%%%%%%%%%%%%%%%%%%%
\subsection*{$\bullet$ Truncation of the OPE}
%%%%%%%%%%%%%%%%%%%%%%%%%%%%%%%%%%%%%%%%%%%
%%%%%%%%%%%%%%%%%%%%%%%%%%%%%%%%%%%%%%%%%%%
\nin In the analysis, we truncate the OPE at the $D=6$ dimension vacuum condensates and assume that the next non-calculated term is of the form:
\beq
\Delta OPE=(\Lambda^2 \tau)\times (D=6~~ \mbox{contributions}).
\eeq
We shall use for 3 flavours: $\Lambda=332(8)\MeV$ using the central value of $\alpha_s(M_Z)=0.1181$ \cite{ANRls}.
%%%%%%%%%%%%%%%%%%%%%%%%%%%%%%%%%%%%%%%%%%%
%%%%%%%%%%%%%%%%%%%%%%%%%%%%%%%%%%%%%%%%%%%
\subsection*{$\bullet$ Final results}
%%%%%%%%%%%%%%%%%%%%%%%%%%%%%%%%%%%%%%%%%%%
%%%%%%%%%%%%%%%%%%%%%%%%%%%%%%%%%%%%%%%%%%%
\nin The final and conservative results are extracted at N3LO assuming a violation of the factorization assumption. Considering the previous estimated contributions as another sources of the errors, we obtain the final results within a Narrow Width Approximation (NWA):
\beq
%\hspace*{-0.1cm}
M_{\pi^+\pi^-} = 1017(159)\MeV,~f_{\pi^+\pi^-}= 1657(277)\keV.
\label{eq:resmol}
\eeq
The different sources of the errors are quoted in Table\,\ref{tab:resqqmol}.
%%%%%%%%%%%%%%%%%%%%%%%%%%%%%%%%%%%%%%%%%%%
%%%%%%%%%%%%%%%%%%%%%%%%%%%%%%%%%%%%%%%%%%%
\subsection*{$\bullet$ Finite width corrections}
%%%%%%%%%%%%%%%%%%%%%%%%%%%%%%%%%%%%%%%%%%%
%%%%%%%%%%%%%%%%%%%%%%%%%%%%%%%%%%%%%%%%%%%
\nin The previous analysis are done using a Narrow Width Approximation (NWA). In order to study the effect of the width, we use the ratio:
\beq
\left( M^{BW}_{\pi^+\pi^-}\right)^2=\frac{\int_{0}^{t_c}~dt~t^5~e^{-t\tau}BW(t)}{\int_{0}^{t_c}~dt~t^4~e^{-t\tau}BW(t)},
\eeq
where, 
\beq
BW(t)=\frac{M_{\pi\pi}\Gamma_{\pi\pi}}{(t-M^{2}_{\pi\pi})^2+M^{2}_{\pi\pi}\Gamma^{2}_{\pi\pi}}
\eeq
and $M_{\pi\pi}$ is the mass from the NWA obtained in Eq.\,\ref{eq:resmol}.

\begin{figure}[hbt]
\begin{center}
\includegraphics[width=7.0cm,height=4.3cm]{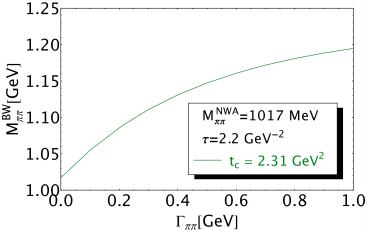}
%\vspace*{-0.5cm}
\caption{\footnotesize The Finite width effect on the mass for $t_c=2.31\,\GeV^2$ corresponding to the central value of $M_{\pi^+\pi^-}=1017\,\MeV$ in a NWA.} 
\label{fig:pp-bw}
\end{center}
\end{figure}

\nin As one can notice from Fig.\,\ref{fig:pp-bw}, varying the value of $\Gamma_{\pi\pi}$ from $0$ to $700\MeV$ (width of the on-shell mass), the width increases the mass by $154\,\MeV$ leading to the estimation:
\beq
M^{BW}_{\pi^+\pi^-}=1171(159)\,\MeV.
\eeq
%%%%%%%%%%%%%%%%%%%%%%%%%%%%%%%%%%%%%%%%%
%%%%%%%%%%%%%%%%%%%%%%%%%%%%%%%%%%%%%%%%%
%%%%%%%%%%%%%%%%%%%%%%%%%%%%%%%%%%%%%%%%%%%
%%%%%%%%%%%%%%%%%%%%%%%%%%%%%%%%%%%%%%%%%%%
%\section{Comments on $\bar{q}q$ and molecule states}
%%%%%%%%%%%%%%%%%%%%%%%%%%%%%%%%%%%%%%%%%%%
%%%%%%%%%%%%%%%%%%%%%%%%%%%%%%%%%%%%%%%%%%%
%\nin -- The masses of the molecule states are about $(230-250)\MeV$ lower than the corresponding $\bar{q}q$ states. However, with the width corrections which act with opposite signs in the two cases, the two estimations tend to meet around $1.1\,\GeV$.\\
%-- The predicted masses of the $\bar{q}q$ meson and molecule states coincide with the one of the lightest scalar gluonium\,\cite{SNA}.\\
%-- At this level, one cannot yet distinguish the $\bar{q}q$, $\pi^+\pi^-$ molecule and gluonium nature of the $\sigma$.
%%%%%%%%%%%%%%%%%%%%%%%%%%%%%%%%%%%%%%%%%%%
%%%%%%%%%%%%%%%%%%%%%%%%%%%%%%%%%%%%%%%%%%%
\section{The tetraquark $\overline{ud}ud\,,\overline{ud}us\,,\overline{us}ds$ states}
%%%%%%%%%%%%%%%%%%%%%%%%%%%%%%%%%%%%%%%%%%%
%%%%%%%%%%%%%%%%%%%%%%%%%%%%%%%%%%%%%%%%%%%
The tetraquark states can be described by many diquark-antidiquark configurations\,\cite{JAFFE,BNNB,CHZ,CS}. For our concern, we choose to work with the configurations "Scalar $\oplus$ Pseudoscalar" and "Vector $\oplus$ Axial-vector":
\bea
\hspace*{-0.7cm}
{\cal O}^{S/P}&=&\epsilon_{abc}\epsilon_{dec}\left[\left(\bar{u}_a\gamma_5\,C\,\bar{q}_{b}^{T}\right)\otimes \left(q'^{T}_{d}\,C\,\gamma_5\,q_e\right)\right.\nnb\\
&&+r\left.\left(\bar{u}_a\,C\,\bar{q}^{T}_b\right)\otimes\left(q'^{T}_{d}\,C\,q_e\right)\right],\nnb
\eea
\bea
\hspace*{-0.7cm}
{\cal O}^{V/A}&=&\hspace*{-0.25cm}\frac{1}{\sqrt{2}}\left[\left(\bar{u}_a\gamma_{\mu}\gamma_5 C \bar{q}^{T}_{b}\right)\left(q'^{T}_{a}C\gamma^{\mu}\gamma_5 q_b-q'^{T}_{b} C \gamma^{\mu}\gamma_5 q_a \right) \right.\nnb\\
&+&\hspace*{-0.3cm}r\left.\left(\bar{u}_a\gamma_{\mu} C \bar{q}^{T}_{b}\right)\left(q'^{T}_{a} C \gamma^{\mu} q_b+q'^{T}_{b} C \gamma^{\mu} q_a \right)\right],
\eea
where, $q\equiv d,s$ and $q'\equiv u,d$; $r$ is an arbitrary mixing parameter.

\nin The analysis is similar to the previous case of molecule states. The final results for the set $r=0,1/\sqrt{2},1$ are given in Tables\,\ref{tab:ressp} and \ref{tab:resva}.
%%%%%%%%%%%%%%%%%%%%%%%%%%%%%%%%%%%%%%%%%%%
%%%%%%%%%%%%%%%%%%%%%%%%%%%%%%%%%%%%%%%%%%%
\section{Nearby radial excitations effects}
\begin{figure}[hbt]
\begin{center}
\includegraphics[width=6.0cm]{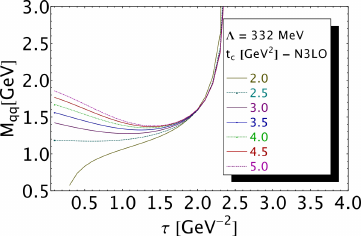}
\includegraphics[width=6.0cm]{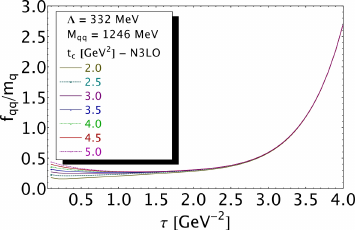}
%\vspace*{-0.5cm}
\caption{\footnotesize The $\tau-$behaviour of the mass and coupling of the $\bar{u}d$ ground state within a "two resonance $\oplus$ QCD continuum" versus the LSR variable $\tau$ and for different values of $t_c$.} 
\label{fig:radeffct}
\end{center}
\end{figure}

%%%%%%%%%%%%%%%%%%%%%%%%%%%%%%%%%%%%%%%%%%%
%%%%%%%%%%%%%%%%%%%%%%%%%%%%%%%%%%%%%%%%%%%
\nin We extend the analysis to the case of the first radial excitations in $SU(2)_F$. In so doing, we shall subtract the contribution of the lowest ground states obtained previously and work with the one resonance parametrization in a higher range of $t_c$. The final results are compiled in Table\,\ref{tab:resrad}.
In order to study the effect of the first radial excitation on the estimation of the lowest ground state mass and coupling, we shall work with the ansatz "two resonances $\oplus$ QCD continuum". As the analysis of the different states will be performed using the same methods, we shall illustrate it in the case of $\bar{u}d$ meson (Fig.\,\ref{fig:radeffct}). We obtain within a NWA:
\beq
\hspace*{-0.1cm}
M_{\bar{u}d} = 1271(124)\MeV,~f_{\bar{u}d}/\overline{m}_{q}= 243(43)\times 10^{-3}.
\eeq
One can notice that these values agree within the errors with the ones in Table\,\ref{tab:resqqmol}. The effect of the nearby first radial excitation on the determination of the lowest ground state mass from the "one resonance $\oplus$ QCD continuum" parametrization is tiny.
%%%%%%%%%%%%%%%%%%%%%%%%%%%%%%%%%%%%%%%%%%%
%%%%%%%%%%%%%%%%%%%%%%%%%%%%%%%%%%%%%%%%%%%
%%%%%%%%%%%%%%%%%%%%%%%%%%%%%%%%%%%%%%%%%%%
\section{Comments and Conclusion}
%%%%%%%%%%%%%%%%%%%%%%%%%%%%%%%%%%%%%%%%%%%
%%%%%%%%%%%%%%%%%%%%%%%%%%%%%%%%%%%%%%%%%%%
We have presented new results for $\pi^+\pi^-$-like molecule states and improved some existing predictions of the masses and couplings of the ordinary $\bar{q}q$ and $(\overline{qq'})(qq')$ tetraquark states. The results including the different sources of the errors are given in Tables\,\ref{tab:resqqmol} to \ref{tab:resrad}.\\
$\circ$ The effects of the SU(3) breaking are only about some few tens of $\MeV$ for the different states.\\
$\circ ~~\sigma/f_0(500)$\\
-- The $R_{P/C}$ condition favours the estimation of the mass of about $1\,\GeV$ but exclude the values around $(0.5-0.6)\MeV$ obtained in the recent literature\,\cite{CHZ,CS}. Our estimate coincides with the on-shell or Breit-Wigner mass expected in Ref.\,\cite{MNO}.\\
-- The constraint on the $\sigma \overline{K}K$ coupling\,\cite{KMSN,MSNW} does not favour the pure $\bar{u}\bar{d}ud$ and $\pi^+\pi^-$ configuration for the $\sigma$. Instead, the observed almost equal  coupling of the $\sigma$ to $pi^+\pi^-$ and $K^+K^-$\,\cite{KMSN,MSNW}  its $\gamma\gamma$ coupling\,\cite{MNO} support its large gluon component\,\cite{SNV, SNB4,SNA}.
\\
$\circ ~~f_0(980)$\\
-- The $\sigma$ and $f_0(980)$ seems to emerge from a maximal meson-gluonium mixing with $(M_{\bar{q}q},\Gamma_{\pi\pi})=(1229,120)\,\MeV$\,\cite{ANRls} and the light scalar gluonium mass $(M_G,\Gamma_G)=(1070,890)\MeV$\,\cite{SNA}.\\
-- The mass of the $K^+K^-$ molecule and the mean prediction of the four-quark states $M_{\bar{u}\bar{s}ds}=1045(112)\MeV$ are compatible with the $f_0(980)$.\\
$\circ ~~a_0(980)$\\
The $\eta\pi^0$ molecule mass and the mean of the four-quark states ones are compatible.\\
$\circ ~~f_0(1370)$\\
This state can be explained by the 1st radial excitation of the $\bar{q}q$ which can mix with the scalar gluonium $M_{\sigma'}=1110(117)\MeV$ to give the large $\pi\pi$ width\,\cite{SNV,SNB4,SNA}. It can also be explained by the mixing of the radial excitation of the four-quark $M_{\bar{u}\bar{d}ud}=1409(112)\MeV$ with the previous gluonium state.\\
$\circ ~~a_0(1450)$\\
This state is compatible with the 1st radial excitation of the $\bar{u}\bar{s}ud$ four-quark which should be almost degenerated to the $\bar{u}\bar{d}ud$ state.\\
$\circ ~~f_0(1500)$\\
It is expected to be a gluonium state from its mass $M_{G'}=1563(141)\MeV$ and from its $U(1)-$like decays $(\eta'\eta, \eta\eta)$\cite{SNV,SNB4,SNA}. From our results, this state is reproduced by the 1st radial excitation of the four-quark state $M_{\bar{u}\bar{d}ud}=1409(112)\MeV$ which may mix with the previous gluonium state.\\
$\circ ~~f_0(1710)$\\
This state can be reproduced by the 1st radial excitation of the four-quark or/and molecule states.\\
$\circ ~~K^{*}_0(700)$\\
The Breit-Wigner mass of $(840\pm 17)\MeV$ \cite{ANRls} is comparable to the ones of the four-quark and the $K\pi$ molecule.\\
$\circ ~~K^{*}_0(1430)$\\
This state can be reproduced by the ordinary $\bar{u}s$ or/and its 1st radial excitation expected to be around $1400\MeV$.

We conclude from this complete analysis of the mass spectrum that the assignement of the nature of the scalar mesons is not crystal clear and needs further studies. Instead, the measurements of the  hadronic and $\gamma\gamma$ couplings as studied in Refs.\,\cite{MNO} and\,\cite{KMSN,MSNW}
help to identify their nature.

%%%%%%%%%%%%%%%%%%%%%%%%%%%%%%
%%%%%%%%%%%%%%%%%%%%%%%%%%%%%%
\begin{table}[H]
\begin{center}
\caption{\scriptsize Sources of errors and values of the masses and couplings of the molecules and ordinary $\bar{q}q$ mesons within NWA. We take $\vert \Delta \tau \vert \simeq 0.2$. In the case of asymmetric errors, we take the mean value.}
%\vspace*{-0.2cm}
\includegraphics[width=16.5cm]{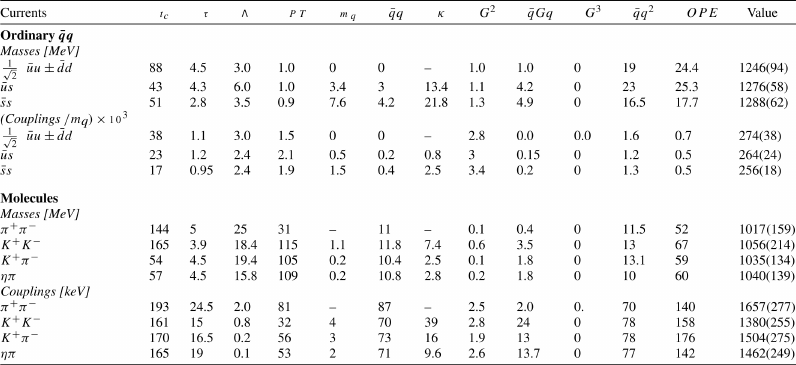}
\label{tab:resqqmol}
\end{center}
\end{table}
%\end{figure*} 
%%%%%%%%%%%%%%%%%%%%%%%%%%%%%%
%%%%%%%%%%%%%%%%%%%%%%%%%%%%%%
%%%%%%%%%%%%%%%%%%%%%%%%%%%%%%
%%%%%%%%%%%%%%%%%%%%%%%%%%%%%%
\begin{table}[H]
\begin{center}
\caption{\scriptsize The same caption as for Table\,\ref{tab:resqqmol} but for the tetraquark states in the Scalar $\oplus$ Pseudoscalar configurations and for three typical values of $r=1,\,1/\sqrt{2},\,0$.}
\vspace*{-0.5cm}
\includegraphics[width=16.5cm,height=9.8cm]{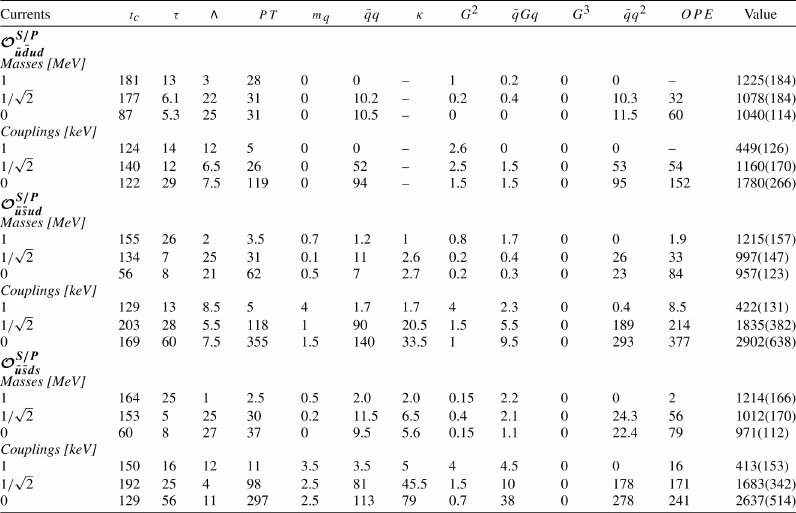}
\label{tab:ressp}
\end{center}
\end{table}
%\end{figure*} 
%%%%%%%%%%%%%%%%%%%%%%%%%%%%%%
%%%%%%%%%%%%%%%%%%%%%%%%%%%%%%
%%%%%%%%%%%%%%%%%%%%%%%%%%%%%%
%%%%%%%%%%%%%%%%%%%%%%%%%%%%%%
\vspace*{-1cm}
\begin{table}[H]
\begin{center}
\caption{\scriptsize The same caption as for Table\,\ref{tab:resqqmol} but for the tetraquark states in the Vector $\oplus$ Axial-vector configurations and for two typical values of $r=1,\,1/\sqrt{2}$.}
%\vspace*{-0.2cm}
\includegraphics[width=16.5cm]{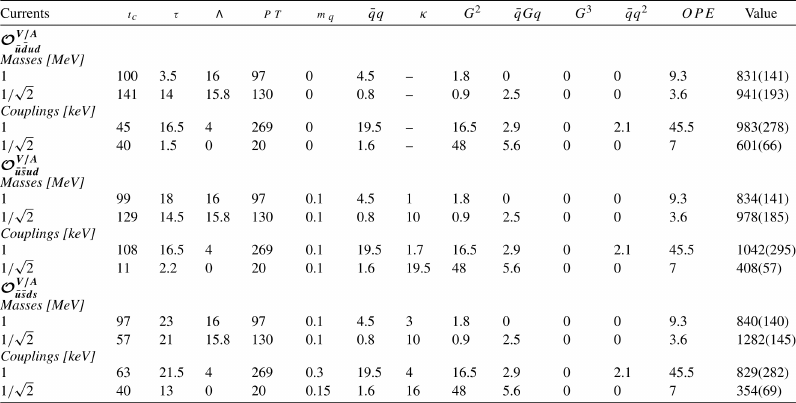}
\label{tab:resva}
\end{center}
\end{table}
%\end{figure*} 
%%%%%%%%%%%%%%%%%%%%%%%%%%%%%%
%%%%%%%%%%%%%%%%%%%%%%%%%%%%%%
%%%%%%%%%%%%%%%%%%%%%%%%%%%%%%
%%%%%%%%%%%%%%%%%%%%%%%%%%%%%%
\vspace*{-1cm}
\begin{table}[H]
\begin{center}
\caption{\scriptsize The same caption as for Table\,\ref{tab:resqqmol} but for the $1$st radial excitations of the different assignments. $r=1,\,1/\sqrt{2},\,0$ are typical values of the four-quark mixing of currents.}
%\vspace*{-0.2cm}
\includegraphics[width=16.5cm]{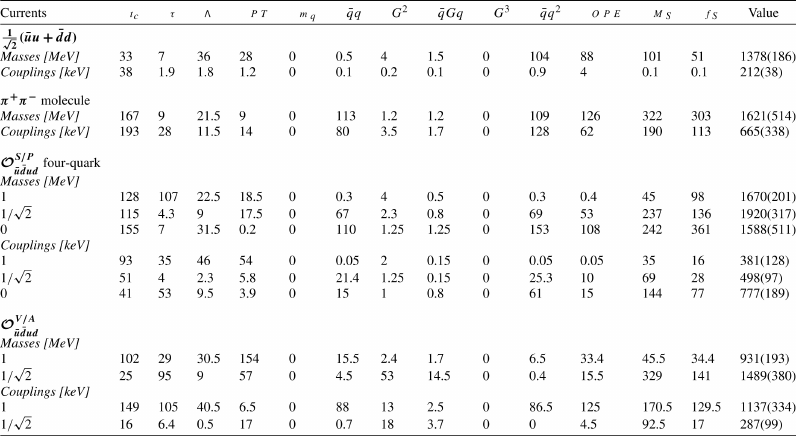}
\label{tab:resrad}
\end{center}
\end{table}
%\end{figure*} 
%%%%%%%%%%%%%%%%%%%%%%%%%%%%%%
%%%%%%%%%%%%%%%%%%%%%%%%%%%%%%
\vfill\eject
 %%%%%%%%%%%%%%%%%%%%%%%%%%%%%


\begin{thebibliography}{999}
 %%%%%%%%%%%%%%%%%%%%%%%%%%%%%
 %%%%%%%%%%%%%%%%%%%%%%%%%%%%%
%%%%%%% LIGHT SCALAR %%%%%%%%
\bibitem{ANRls} R.M. Albuquerque, S. Narison and D. Rabetiarivony, {\it Nucl. Phys.} {\bf A1039}, 122743 (2023).
\bibitem{SNB4} S. Narison, {\it Nucl. Phys.} {\bf B509} (1998) 312.
\bibitem{SNA} S. Narison, {\it Nucl. Phys.} {\bf A1017} (2022) 122337.
\bibitem{MNP} G. Mennessier, S. Narison, N. Paver, {\it Phys. Lett.} {\bf B158} (1985) 153-157.
\bibitem{NPRT} S. Narison, N. Paver, E. de Rafael, D. Treleani, {\it Nucl. Phys.} {\bf B212} (1983) 365.
\bibitem{SND} S. Narison, {\it Phys. Rev.} {\bf D73} (2006) 114024.
\bibitem{RRY} L.J. Reinders, H. Rubinstein, S. Yazaki, {\it Phys. Rep.} {\bf 127} (1985)1.
\bibitem{SNB5} S. Narison, {\it Phys. Lett.} {\bf B175} (1986) 88.
\bibitem{BSN} A. Bramon, S. Narison, {\it Mod. Phys. Lett.} {\bf A4} (1989) 1113.
\bibitem{SNV} S. Narison, G. Veneziano, {\it Int. J. Mod. Phys.} {\bf A4} (11) (1989) 2751.
%%%%%%%%%%%  FOUR-QUARK %%%%%%%%%%%%%%%%%%%%%%
\bibitem{JAFFE} R.L. Jaffe, {\it Phys. Rev.} {\bf D15} (1977) 267; 281; {\it Phys. Rep.} {\bf 409} (2005) 1.
\bibitem{ADS} N.N. Achasov, S.A. Devyanin, G.N. Shestakov, {\it Z. Phys.} {\bf C16} (1984) 55.
\bibitem{IW} N. Isgur, J. Weinstein, {\it Phys. Rev.} {\bf D41} (1990) 2236.
\bibitem{LP} J.I. Latorre, P. Pascual, {\it J. Phys.} {\bf G11} (1985) L231.
\bibitem{BNNB} T.V. Brito, F.S. Navarra, M. Nielsen, M.E. Bracco, {\it Phys. Lett.} {\bf B608} (2005) 69.
\bibitem{CHZ} H.X. Chen, A. Hosaka, S.L. Zhu, {\it Phys. Rev.} {\bf D76} (2007) 094025.
\bibitem{CS} B.A. Cid-Mora, T.G. Steele, {\it Nucl. Phys.} {\bf A1028} (2022) 122538.
%%%%%%%% LSR %%%%%%%%
\bibitem{BELL} J.S. Bell and R.A. Bertlmann, {\it Nucl. Phys.} {\bf B177} (1981) 218.
\bibitem{BELLa} J.S. Bell and R.A. Bertlmann, {\it Nucl. Phys.} {\bf B187} (1981) 285.
\bibitem{BNR} C. Becchi et {\it al.}, {\it Z. Phys.} {\bf C8}, 335 (1981).
\bibitem{BERT} R.A. Bertlmann, {\it Acta Phys. Austriaca} {\bf 53}, 305 (1981)% and references therein.
\bibitem{NEUF} R.A. Bertlmann and H. Neufeld, {\it Z. Phys.} {\bf C27} (1985)  437.
\bibitem{SNR} S. Narison and E. de Rafael,  {\it Phys. Lett.} {\bf B522} (1986) 266.
\bibitem{SNB1} S. Narison, {\it QCD as a theory of hadrons, Cambridge Monogr. Part. Phys. Nucl. Phys. Cosmol.} {\bf 17} (2002) 1.%; [hep-ph/0205006].
\bibitem{SNB2} S. Narison, {\it QCD Spectral Sum Rules,  World Sci. Lect. Notes Phys.,} vol. {\bf 26}, ISBN 9780521037310, 1989, p. 1.
%%%%%%% QSSR %%%%%%%
\bibitem{SVZa} M.A. Shifman, A.I. Vainshtein and V.I. Zakharov, {\it Nucl. Phys.} {\bf B147} (1979) 385, 448.
\bibitem{Za} V.I. Zakharov, {\it Int. J. Mod. Phys.} {\bf A14}, 4865 (1999).
\bibitem{SNqcd22} S. Narison, {\it Nucl. Part. Phys. Proc.} 324-329, (2023) 94-106.
%\bibitem{SNB1} S. Narison, {\it QCD as a theory of hadrons, Cambridge Monogr. Part. Phys. Nucl. Phys. Cosmol.} {\bf 17} (2002) 1; [hep-ph/0205006].
%\bibitem{SNB2}S. Narison, {\it QCD Spectral Sum Rules,  World Sci. Lect. Notes Phys.,} vol. 26, ISBN 9780521037310, 1989, p. 1.
\bibitem{SNB3}S. Narison, {\it Phys. Rept.}  {\bf 84} (1982) 263; {\it Acta Phys. Pol.} {\bf B26} (1995) 687. 
\bibitem{CK} E. de Rafael, hep-ph/9802448.
\bibitem{YND} F. J. Yndurain, {\it The Theory of Quark and Gluon Interactions,} 3rd ed. (Springer, New York, 1999).
\bibitem{PAS} P. Pascual and R. Tarrach, {\it QCD: Renormalization for Practitioner} (Springer, New York, 1985).
%\bibitem{RRY} L. J. Reinders, H. Rubinstein, and S. Yazaki, {\it Phys. Rep.} 127, 1(1985).
\bibitem{IOFF} B. L. Ioffe, {\it Prog. Part. Nucl. Phys.} {\bf 56}, 232(2006).
\bibitem{DOSCH} H. G. Dosch, {\it Non-Perturbative Methods}, edited by S. Narison (World Scientific, Singapor,1985).%%%%%%% OUR WORK %%%%%%%%
\bibitem{LNSR} S. Li, S. Narison, T.G. Steele and D. Rabetiarivony, {\it Phys. Lett.} {\bf B849} (2024) 138454.
%\bibitem{ANRls} R. M. Albuquerque, S. Narison and D. Rabetiarivony, {\it Nucl. Phys.} {\bf A 1039} (2023) 122743.
\bibitem{ANRTm} R. M. Albuquerque, S. Narison and D. Rabetiarivony, {\it Nucl. Phys.} {\bf A1034} (2023) 122637.
\bibitem{ANR21} R. M. Albuquerque, S. Narison and D. Rabetiarivony, {\it Phys. Rev.} {\bf D103}, 074015 (2021); {\it Phys. Rev.} {\bf D105}, 114035 (2022).
\bibitem{ANR22pa} S. Narison and D. Rabetiarivony, {\it Nucl. Part. Phys. Proc.} 324-329, (2023) 54-58.
\bibitem{ANR22p} S. Narison and D. Rabetiarivony, {\it Nucl. Part. Phys. Proc.} 324-329, (2023) 49-53.
\bibitem{NR1} S. Narison and D. Rabetiarivony, {\it Nucl. Part. Phys. Proc.} 318-323, (2022) 96-101.
\bibitem{ANR1} R. Albuquerque et {\it al.}, {\it Nucl. Phys.} {\bf A1007} (2021) 122113.
\bibitem{ANRR2} R. M. Albuquerque et {\it al.}, {\it Nucl. Part. Phys. Proc.} 312-317, (2021) 120-124.
\bibitem{ANRR1a} R. M. Albuquerque et {\it al.}, {\it Nucl. Part. Phys. Proc.} 300-302, (2018) 186-195.
\bibitem{SU3} R. Albuquerque et {\it al.}, {\it Int. J. Mod. Phys.} {\bf A33} (2018), 1850082.
\bibitem{ANRR1} R. M. Albuquerque et {\it al.}, {\it Nucl. Part. Phys. Proc.} 282-284, (2017) 83.
\bibitem{SNX1} R. Albuquerque et {\it al.}, {\it Int. J. Mod. Phys.} {\bf A31} (2016) no.17, 1650093.
\bibitem{SNX2} R. Albuquerque et {\it al.}, {\it Int. J. Mod. Phys.} {\bf A31} (2016) no. 36, 1650196.
%\bibitem{AFNR} R. Albuquerque et {\it al.}, {\it Phys. Lett.} {\bf B 175}, (2012) 129.
%%%%%%%%%%%%%%%%%%%%
\bibitem{SND} S. Narison, {\it Phys. Rev.} {\bf D73} (2006) 114024.
\bibitem{JM} M. Jamin, M. Munz, {\it Z. Phys.} {\bf C66} (1995) 633.
\bibitem{SNB6} S. Narison, {\it Phys. Lett.} {\bf B738} (2014) 346.
\bibitem{SNZ} S. Narison, V.I. Zakharov, {\it Phys. Lett.} {\bf B522} (2001) 266.
\bibitem{ZAK} V.I. Zakharov, {\it Nucl. Phys. B, Proc. Suppl.} 164 (2007) 240.
\bibitem{SNBs} S. Narison, {\it Nucl. Phys. B, Proc. Suppl.} 164 (2007) 225.
\bibitem{ANDREEV} O. Andreev, {\it Phys. Rev.} {\bf D73} (2006) 107901.
\bibitem{AZAK} O. Andreev, V.I. Zakharov, {\it Phys. Rev.} {\bf D74} (2006) 025023; {\it Phys. Rev.} {\bf D76} (2007) 047705.
\bibitem{JSNR} F. Jugeau, S. Narison, H. Ratsimbarison, {\it Phys. Lett.} {\bf B722} (2013) 111.
\bibitem{CSNZ} K.G. Chetyrkin, S. Narison, V.I. Zakharov, {\it Nucl. Phys.} {\bf B550} (1999) 353.
\bibitem{KMSN} R. Kaminski, G. Mennessier, S. Narison, {\it Phys. Lett.} {\bf B680} (2009) 148.
\bibitem{MSNW} G. Mennessier, S. Narison, X.G. Wang, {\it Phys. Lett.} {\bf B696} (2011) 40.
\bibitem{MNO} G. Mennessier, S. Narison, W. Ochs, {\it Phys. Lett.} {\bf B665} (2008) 205.
 %%%%%%%%%%%%%%%%%%%%%%%%%%%%%
\end{thebibliography}
\end{document}